\documentclass[useAMS,usenatbib]{mn2e}
\usepackage{epsfig}

\newcommand{\be}{\begin{eqnarray}}
\newcommand{\ee}{\end{eqnarray}}
\def\lsim{\,\lower2truept\hbox{${< \atop\hbox{\raise4truept\hbox{$\sim$}}}$}\,}
\def\gsim{\,\lower2truept\hbox{${> \atop\hbox{\raise4truept\hbox{$\sim$}}}$}\,}

\title[Detectable Signatures of Cosmic Radiative Feedback]
{Detectable Signatures of Cosmic Radiative Feedback}

\author[R.~Schneider et al.]{R.~Schneider$^{1}$\thanks{raffa@arcetri.astro.it}, R.~Salvaterra$^2$, 
T.~Roy~Choudhury$^3$, A.~Ferrara$^4$, C.~Burigana$^{5}$ \newauthor and L.~A.~Popa$^{5,6}$\\
$^1$ INAF - Osservatorio Astrofisico di Arcetri, Largo Enrico Fermi 5, 50125 Firenze, Italy \\
$^2$ Dipartimento di Fisica G. Occhialini, Universit{\'a} degli Studi di Milano Bicocca, Piazza della Scienza 3, I-20126 Milano, Italy\\
$^3$ Institute of Astronomy, Madingley Road, Cambridge CB3 0HA, UK\\
$^4$ SISSA/International School for Advanced Studies, Via Beirut 4, 34100 Trieste, Italy \\
$^5$ INAF-IASF Bologna, via Gobetti 101, I-40129 Bologna - Italy \\
$^6$ Institute for Space Sciences, Bucharest-Magurele, Str. Atomostilor, 409, PoBox Mg-23,Ro-077125, Romania}

\begin{document}

\date{28 November 2007}
\pagerange{\pageref{firstpage}--\pageref{lastpage}}
\pubyear{2007}

\maketitle
\label{firstpage}

\begin{abstract}
We use a semi-analytical model to study the impact of reionization, and the associated radiative
feedback, on galaxy formation.
Two feedback models have been considered: (i) a standard prescription, according to which
star formation is totally suppressed in galaxies with circular velocity below a 
critical threshold (model CF06) and (ii) a characterization based on the filtering scale (model G00), allowing for a 
gradual reduction of the gas available for star formation in low-mass galaxies. 
In model CF06 reionization starts at $z \lsim 15-20$, is $85$\% complete by $z \sim 10$; 
at the same $z$, the ionized fraction is $16\%$ in model G00. The models match SDSS constraints
on the evolution of the neutral hydrogen fraction at $z < 7$, but predict 
different Thomson optical depths, $\tau_\mathrm{e}=0.1017$ (CF06), and  $0.0631$ (G00); such values are within
1$\sigma$ of the {\it WMAP} 3-yr determination.  Both models are in remarkable good agreement with additional existing 
data (evolution of Lyman-limit systems, cosmic star formation history, high-$z$ galaxy counts, IGM thermal history), 
which therefore cannot be used to discriminate among different feedback models.    
Deviations among radiative feedback prescriptions emerge when considering the expected HI 21 cm background 
signal, where a $\sim 15$~mK absorption feature in the range 75-100 MHz is present in model G00 and a global shift 
of the emission feature preceding reionization towards larger frequencies occurs in the same model. Single dish 
observations with existing or forthcoming low-frequency radio telescopes can achieve mK sensitivity, allowing the
identification of these features provided that foregrounds can be accurately subtracted.
\end{abstract}


\section{Introduction}
\label{intro}

The process of cosmic reionization is of primary importance in the
evolution of the Universe. Reionization is a consequence of the 
formation of the first luminous sources that shine after the Dark Ages,
but it also has a dramatic impact on subsequent galaxy evolution.
Despite the recent progress in observational and theoretical studies,
fundamental questions such as what are the dominant sources of reionization
and how long did the process take still require complete and definite 
answers.

The latest analysis of Ly$\alpha$ absorption in the 
spectra of the 19 highest redshift Sloan Digital Sky Survey (SDSS) quasars (QSOs) 
shows a strong evolution of the Gunn-Peterson Ly$\alpha$ opacity at $z \sim 6$ \citep{Fan2006, Gallerani2006}.
This result, together with the downward revision of the
electron scattering optical depth to $\tau_e = 0.09 \pm 0.03$ in the release of 
the 3-yr {\it Wilkinson Microwave Anisotropy Probe} ({\it WMAP}) data \citep{Page2007,Spergel2007}, 
is consistent with ``minimal reionization models'' which do not require the presence of
very massive ($M>100 M_\odot$) Population III stars \citep{Choudhury2006, Gnedin2006}. 

Choudhury \& Ferrara (2005, 2006) have developed
a self-consistent formalism which allows to jointly study cosmic
reionization and the thermal history of the intergalactic medium 
(IGM). Their semi-analytic model accounts for inhomogeneous IGM 
density distribution, three different classes of ionizing photons
(Population III, Population II stars and QSOs), and radiative 
feedback inhibiting star formation in low-mass galaxies. Note that Population III (Pop III) 
stars in this model are characterized by a standard Salpeter Initial Mass Function (IMF) 
extending in the range $1-100~M_\odot$. 
The model free parameters are constrained by a wide range of observational
data. From this analysis, it emerges that hydrogen reionization 
starts around $z \approx 15$ driven by the contribution of Pop III 
stars and it is 80\% complete by $z \approx 10$; below this 
redshift, the contribution of Pop III stars decreases because of the combined
action of radiative and chemical feedback. As a consequence, reionization is 
extended considerably, completing only at $z \approx 6$ \citep{Choudhury2006}. 

These results are consistent with the study of \citet{Gnedin2006},
who used a set of numerical simulations with different spatial and 
mass resolutions to make a detailed comparison with SDSS and {\it WMAP}
data. Assuming the sources of ionizing photons to be Population II (Pop II)
stars and QSOs, they find that the simulations can match the SDSS 
observations in the range $5 < z < 6.2$, but do not have enough 
resolution to resolve the earliest stages of star formation. 
Depending on the fractional contribution of Pop III stars, the optical
depth to Thomson scattering varies in the range $0.06 < \tau_e < 0.10$, 
consistent with the 3-yr {\it WMAP} results.

Since it appears that minimal reionization models provide the best 
theoretical frameworks for the existing data, we can use these models
to explore the effects of reionization on galaxy formation, to which  
we will refer to as ``radiative feedback". 

It is well known that the temperature increase of the cosmic gas 
in ionized regions leads to a dramatic suppression of the formation 
of low-mass galaxies. Such suppression has been found to become effective
in dark matter halos with circular velocity below a critical one, $v_\mathrm{crit}$, whose value 
is loosely constrained by a number of studies \citep{Thoul1996, Susa2000, Kitayama2001, Machacek2001, Dijkstra2004}. 
Values in the range $v_\mathrm{crit} = 10-50$ km/s are found, depending on the assumed 
intensity of the UV background, the inclusion of radiative transfer 
effects, the numerical scheme adopted, and the investigated redshift 
range (see \citealt{Ciardi2005}~for an extensive review). This 
radiative feedback prescription is therefore unable to quantify 
the decrease in the amount of gas available for star formation in low-mass halos: 
it simply states that galaxies can form stars unimpeded provided that their halos  have 
circular velocities $> v_\mathrm{crit}$.

A perhaps more complete characterization of radiative    feedback on low-mass
galaxy has been obtained by \citet{Gnedin2000}. Based on cosmological simulations 
of reionization, this study shows that the effect of photoionization is 
controlled by a single mass scale in both the linear and non linear regimes. 
The gas fraction within dark matter halos  at any given moment is fully specified 
by the current filtering mass, which directly corresponds to the length 
scale over which baryonic perturbations are smoothed in linear theory. The 
results of this study provide a quantitative description of radiative    feedback,  
independently of whether this is physically associated to photoevaporative flows 
or due to accretion suppression. 

In the present paper, we show that these two different
feedback prescriptions have important effects on the reionization 
history and on the properties of the sources which dominate the process. 
Using the semi-analytic model of Choudhury \& Ferrara (2005, 2006), we 
compare the results obtained for these two competitive feedback 
models with a wide range of observational data, ranging from 
the redshift evolution of Lyman-limit absorption systems, 
the Gunn-Peterson and electron scattering optical depths, the cosmic star 
formation history, and number counts of high-$z$ sources in the NICMOS 
Hubble Ultra Deep Field (HUDF). We find that in spite of this demanding 
benchmark of observations, existing data are unable to discriminate 
among the two reionization histories.
We therefore explore an alternative method to break these degeneracies 
using future 21 cm experiments such as the Low Frequency Array (LOFAR),
the 21-Centimeter Array (21CMA), the Mileura Wide Field Array (MWA), and
the Square Kilometer Array (SKA); these instruments are expected to reach 
the sensitivity required to perform accurate maps of the neutral hydrogen 
distribution from the Dark Ages to the latest stages of the reionization
process.

The paper is organized as follows: in Section \ref{models} we give a brief
description of the semi-analytic formalism and present the results of the
two radiative    feedback models, comparing these with existing observations;
in Section \ref{21cm} we compute the predicted global 21 cm background in the
two models and discuss its detectability with on-going and future facilities.
Finally, Section \ref{summary} summarizes our conclusions.
 
Throughout this paper, we adopt a flat $\Lambda$CDM cosmological model consistent with 
3-yr {\it WMAP} data (Spergel et al. 2007), with matter and cosmological constant 
density parameters $\Omega_m=0.24$ and $\Omega_\Lambda=0.76$, reduced Hubble constant 
$h=0.73$, baryon density $\Omega_b h^2=0.022$, density contrast $\sigma_8=0.74$, and 
adiabatic scalar perturbations (without running) with spectral index $n_s=0.95$. 
We also assume a Cosmic Microwave Background (CMB) temperature of 2.725~K \citep{mather99}.

\begin{table*}
\begin{minipage}{120mm}
\caption{Best-fit parameters for models CF06 and G00 shown in Figs.~\ref{fig:modelCF06} and \ref{fig:modelG00}. 
For the same models, we also report the residual volume-averaged neutral hydrogen fraction at redshift 6, $x_\mathrm{HI}(6)$, 
the volume-averaged electron fraction at redshfit 10, $x_\mathrm{e}(10)$, and the Thomson scattering optical depth 
$\tau_\mathrm{el}$ (see text).}
\label{table:model}
\centering
\begin{tabular}{lccccccc}
\hline
Model & $\epsilon_\mathrm{*,II}$ & $f_\mathrm{esc, II}$ &  $\epsilon_\mathrm{*,III}$ & $f_\mathrm{esc, III}$ & $x_\mathrm{HI}(6)$ 
& $x_\mathrm{e}(10)$ & $\tau_\mathrm{el}$ \\
\hline
CF06 & 0.10 & 0.01 & 0.03 & 0.68 & $4 \times 10^{-4}$    & 0.85  & 0.1017 \\
G00  & 0.10 & 0.01 & 0.01 & 0.12 & $3.9 \times 10^{-4}$  & 0.16 & 0.0631 \\
\hline
\end{tabular}
\end{minipage}
\end{table*}

\section{Reionization Models}
\setcounter{equation}{0}
\label{models}

In this Section we first summarize the main features of the semi-analytical model developed
by \citet{Choudhury2005} with the additional physics introduced in \citet{Choudhury2006}. 
We then describe the two alternative radiative    feedback prescriptions and compare
the resulting reionization histories with existing data.

\subsection{Model description}

The main features of the formalism developed by Choudhury \& Ferrara (2005, 2006) can be
summarized as follows:

\begin{itemize}

\item Inhomogeneous reionization: the model accounts for IGM inhomogeneities by adopting
the procedure of Miralda-Escud{\'e}, Haehnelt \& Rees (2000). The overdensity distribution
is assumed to be lognormal. The distribution determines the mean free path of photons,

\be
\lambda_\mathrm{mfp}(z)=\frac{ \lambda_0 }{[1-F_V(z)]^{2/3}}
\label{eq:mfp}
\ee
\noindent
where $F_V$ is the volume fraction of ionized regions and $\lambda_0$ is a normalization constant
fixed by comparing with low-redshift observations of Lyman-limit systems. 

\item Sources of ionizing photons: the IGM is treated as a multiphase medium, following the thermal 
and ionization histories of neutral, HII, and HeIII regions simultaneously. Three sources have been 
assumed to contribute to the ionizing flux: (i) Pop III stars, assumed to be distributed according 
to a standard Salpeter IMF, as suggested by the combination of constraints from source counts at 
$z \sim 10$ and {\it WMAP} data \citep{Schneider2006}; (ii) Pop II stars, with $Z=0.2 Z_{\odot}$ 
and Salpeter IMF; (iii) QSOs which are significant sources of hard photons at $z \lsim 6$. 
The emission is modelled using the stellar population 
synthesis templates from \citet{Bruzual2003} for Pop II stars and from \citet{Schaerer2002} for 
Pop III stars.  

\item  Chemical feedback: the transition from Pop III to Pop II stars is controlled by 
chemical feedback and occurs over a prolonged epoch rather than at a precise transition redshift. 
Using the merger-tree model developed by Schneider et al. (2006), at each redshift we classify 
star forming halos as hosting Pop II (Pop III) stars depending on whether the halo itself or 
any of its progenitors have (have not) already experienced an episode of star 
formation\footnote{Since Pop III stars are assumed to be distributed according to a Salpeter IMF, star
formation in Pop III halos always leads to metal-enrichment as a consequence of type-II supernova
explosions. This is equivalent to the strong feedback case, $f_{sn}=1$, in the formalism of
Schneider et al. (2006).}. The fraction of Pop III halos decreases with time, 
being $\approx 0.4, 0.32$, and $0.23$ at $z = 15, 10$, and $5$, respectively. 
At each redshift, Pop III stars are confined to form in halos with masses 
$10^8 - 10^9 M_{\odot}$ which are large enough to form stars but small enough 
to be relatively unpolluted by their progenitors.

\item Escape fractions: to reduce the number of model free parameters, 
Choudhury \& Ferrara (2006) use a physical argument to relate the escape
fractions of Pop III and Pop II ionizing photons. These scale according
to the number of ionizing photons produced and depend on a single free-parameter, 
$N_\mathrm{abs}$, which represents the number of ionizing photons absorbed 
within a star-forming halo, through the relation
\be
\eta_\mathrm{esc} \equiv \frac{N_\mathrm{abs}}{\epsilon_\mathrm{*,II} N_{\gamma,\mathrm{II}}}.
\label{eq:esc}
\ee
\noindent
The escape fractions for Pop II and Pop III stars are then given by the following
expressions,
\be
f_\mathrm{esc, II} = 1 - \mathrm{Min}[1,\eta_\mathrm{esc}]
\ee
\be
f_\mathrm{esc, III} = 1 - \mathrm{Min}\left[1,\frac{\epsilon_\mathrm{*,II} 
N_{\gamma,\mathrm{II}}}{\epsilon_\mathrm{*,III} N_{\gamma,\mathrm{III}}}\, \eta_\mathrm{esc}\right];
\ee
\noindent
where $N_{\gamma,\mathrm{II}}$ ($N_{\gamma,\mathrm{III}}$) are the number of photons produced by 
Pop II (Pop III) stars per unit mass of stars formed, and $\epsilon_\mathrm{*,II}$ ($\epsilon_\mathrm{*,III}$)
are the Pop II (Pop III) star formation efficiencies.
\end{itemize}

Further details on the models can be found in the original papers (Choudhury \& Ferrara 2005, 2006).

\subsection{Radiative feedback}

A variety of feedback mechanisms can suppress star formation in mini-halos, i.e. halos with 
virial temperatures $< 10^4$~K, particularly if their clustering is 
taken into account \citep{Kramer2006}. We therefore assume that stars can form in halos
down to a virial temperature of $10^4$~K, consistent with the interpretation of the 3-yr 
{\it WMAP} data (Haiman \& Bryan 2006; but see also Alvarez et al. 2006). We then assume that
these halos  can be affected by radiative    feedback according to two alternative prescriptions:

\begin{enumerate}

\item Following \citet{Choudhury2006}, we assume that in photoionized regions halos  can form stars 
{\it only} if their circular velocity exceeds the critical value 

\be
v_\mathrm{crit} = \frac{2 k_B T}{\mu m_\mathrm{p}},  
\ee 

\noindent
where $\mu$ is the mean molecular weight, $m_\mathrm{p}$ is the proton mass, and $T$ is the average temperature of 
ionized regions, which we can compute self-consistently from the multiphase IGM model. Typically, for a 
temperature of $3 \times 10^4$~K, the minimum circular velocity is $\approx 30$~km/s, within the range of
values found in the literature (see Section \ref{intro} and \citealt{Ciardi2005} for a complete reference list).  
Note that the above value for $v_\mathrm{crit}$ evolves according to the gas temperature and it is not fixed to 
a particular value. Hereafter, we shall refer to this model as CF06.

\item Following \citet{Gnedin2000}, we assume that the average baryonic mass $M_{b}$ within halos  in photoionized regions 
is a fraction of the universal value $f_{b} = \Omega_b/\Omega_m$, given by the following fitting formula,

\be  
\frac{M_b}{M} = \frac{f_b}{[1+(2^{1/3}-1) M_\mathrm{C}/M]^3},
\ee
\noindent
where $M$ is the total halo mass, and $M_\mathrm{C}$ is the total mass of halos  that on average retain 50\% of their gas mass.
A good approximation for the characteristic mass $M_\mathrm{C}$ is given by the linear-theory filtering mass,

\be
M_\mathrm{F}^{2/3} = \frac{3}{a} \int_0^a da^\prime M_\mathrm{J}^{2/3}(a^\prime) 
\left[1-\left(\frac{a^\prime}{a}\right)^{1/2}\right],
\ee
\noindent
where $a$ is the cosmic scale factor, 

\[
M_\mathrm{J} \equiv \frac{4 \pi}{3} \bar{\rho} \left(\frac{\pi c_\mathrm{s}^2}{G \bar{\rho}}\right)^{3/2},
\]
\noindent
is the Jeans mass, $\bar{\rho}$ is the average total mass density of the Universe, and $c_\mathrm{s}$ is the gas sound speed. 
Since the gas density and temperature are not uniform, the sound speed is computed according to

\[
c_\mathrm{s}^2 = \frac{5}{3} \frac{k_B \langle T\rangle_\mathrm{V}}{\mu m_\mathrm{p}},
\]

\noindent
where $\langle T\rangle_\mathrm{V}$ indicates the volume-averaged gas temperature \citep{Gnedin2000}. Hereafter, we will refer to this 
model as G00. 
\end{enumerate}

\begin{figure}
\center{ \epsfig{file=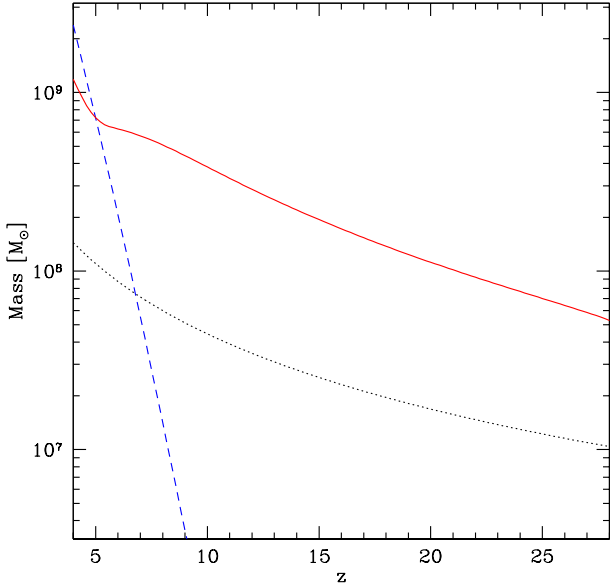, width=8.5cm} }
\caption{Redshift evolution of the filtering mass for the G00 model (dashed line), the minimum mass of star forming halos (dotted line), and the critical mass in ionized 
regions for the CF06 model (solid line).}
\label{fig:masses}
\end{figure}

In Fig.~\ref{fig:masses} the filtering mass for the G00 model (dashed line) and the critical mass in ionized 
regions for the CF06 model (solid line) are plotted as a function of redshift.
While these masses play different roles in the two feedback models, it is interesting to compare them with 
the minimum halo mass allowed to form stars (dotted line).

\begin{figure*}
\center{ \epsfig{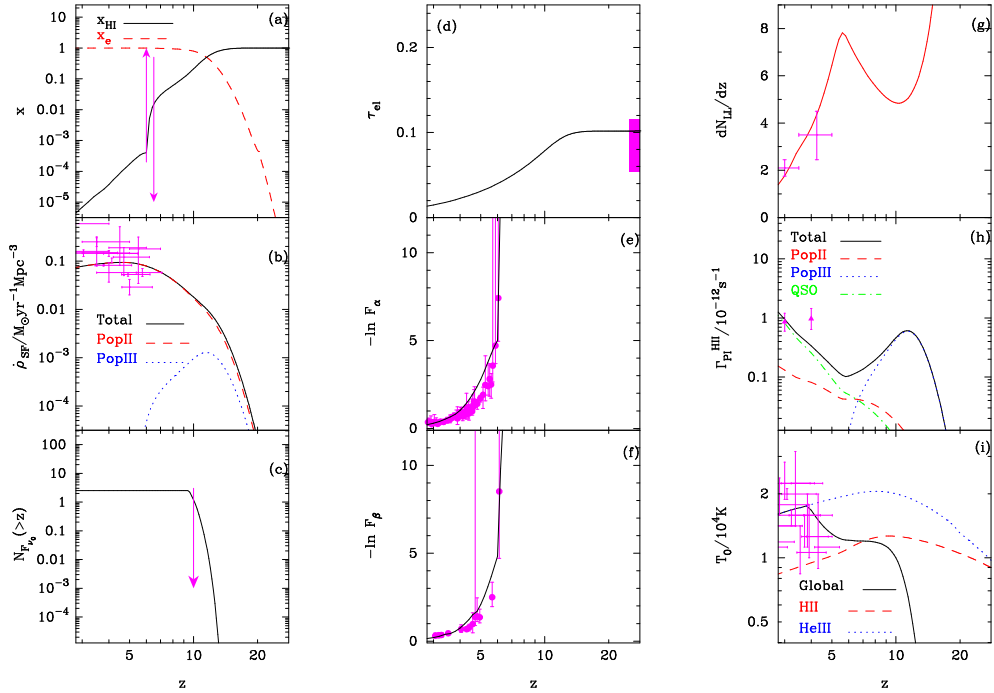} }
\caption{The best-fitting model CF06. The adopted parameters are given in Table~\ref{table:model}. The panels show
the redshift evolution of: (a) the volume-averaged electron and HI fraction. The arrows show an observational lower limit from QSO
absorption lines at $z=6$ and upper limit from Ly$\alpha$ emitters at $z=6.5$; (b) the cosmic star formation history, with
the contribution of Pop III and Pop II stars. Observational data are taken from the compilation of Nagamine et al. (2004); 
(c) the number of source counts above a given redshift, with the observational upper limit from NICMOS HUDF \citep{Bouwens2005};
(d) the electron scattering optical depth, with observational constraints from 3-yr {\it WMAP} data; (e) Ly$\alpha$ effective optical
depth with data from \citet{Songaila2004}; (f) Ly$\beta$ effective optical depth with data from \citet{Songaila2004}; (g) the evolution of
Lyman-limit systems with observational data from \citet{Storrie-Lombardi1994}; (h) photoionization rates for neutral hydrogen, with
estimates from numerical simulations (points with errorbars, \citealt{Bolton2005}); (i) temperature evolution of the mean density IGM,
with observational data from \citet{Schaye1999}.   
}
\label{fig:modelCF06}
\end{figure*}

\subsection{Results}

\begin{figure*}
\center{ \epsfig{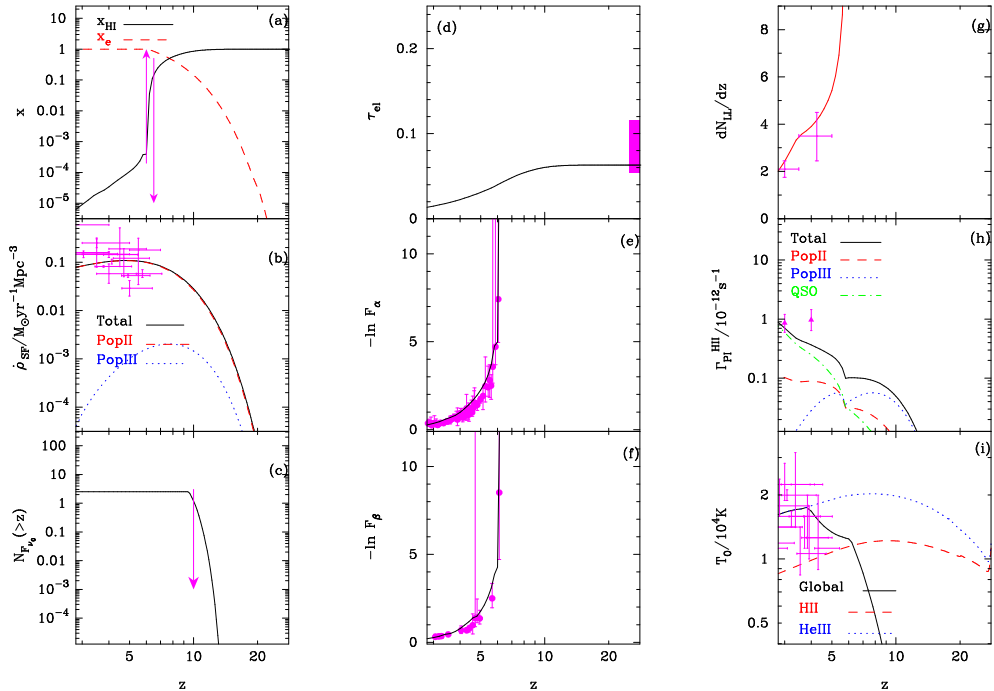} }
\caption{Same as Fig.~\ref{fig:modelCF06} but for model G00. The adopted parameters are shown in Table~\ref{table:model}.
}
\label{fig:modelG00}
\end{figure*}

The semi-analytical model depends only on four free parameters: the star formation efficiencies of Pop II
and Pop III stars, the parameter $\eta_\mathrm{esc}$ which is related to the escape fraction of ionizing 
photons emitted by Pop II and Pop III stars (see equation~\ref{eq:esc}), and the normalization of the photon 
mean free path, $\lambda_0$ (see equation~\ref{eq:mfp}),
which is fixed to reproduce low-redshift observations of Lyman-limit systems \citep{Choudhury2005}. 
In Figs.~\ref{fig:modelCF06} and \ref{fig:modelG00} we show how the two models compare to existing observational 
data. The best-fitting parameters for each model are reported in 
Table~1. Both models show a remarkable agreement with a variety of observations. However, the two feedback 
prescriptions have a noticeable impact on the overall reionization history and the relative contribution
of different ionizing sources. The main reason is that, although the two models predict similar global star formation 
histories dominated by Pop II stars (panels b), the Pop III star formation rates have
markedly different redshift evolution. In fact, chemical feedback forces Pop III stars to live preferentially  
in the smallest, quasi-unpolluted halos (virial temperature $\gsim 10^4$~K), which are 
those most affected by radiative feedback (see Figure \ref{fig:masses}).

In model CF06, where star formation is totally suppressed below $v_\mathrm{crit}$, Pop III 
stars disappear at $z \sim 6$; conversely, in model G00, where halos  suffer a gradual reduction of the available 
gas mass, Pop III stars continue to form at $z \lsim 6$, though with a declining rate. 
As the star formation and photoionization rate at these redshifts are well constrained by observations,
the star formation efficiency and escape fraction of Pop III stars need to be lower in model G00 in
order to match the data (see Table~1). Because of this low  $f_\mathrm{esc, III} \times \epsilon_\mathrm{*,III}$,
reionization starts late in model G00 ($z \lsim 15$) and only 16\% of the volume is reionized at $z=10$
(reionization starts at $z \sim 20$ in model CF06 and it is 85\% complete by $z=10$). 
For $6 < z < 7$, QSOs, Pop II and Pop III give a comparable contribution to the total photo-ionization
rate in model G00, whereas in model CF06 reionization at $z < 7$ is driven primarily by QSOs, with a smaller
contribution from Pop II stars only. The predicted electron scattering optical depths are consistent with 
{\it WMAP} 3-yr data but with $\sim 1 \sigma$ difference among the two models, leaving a chance of 
probing them with forthcoming CMB experiments. Of course, it would still be interesting to
see whether we can find any additional detectable signatures, particularly with respect to the upcoming 
21cm experiments, which is discussed in the next section.

\section{21 cm Signal}
\label{21cm}

The predicted free electron fraction and gas temperature evolution in the redshift 
range $7 < z < 20$ represent the largest difference among the two feedback models.
Therefore, the global 21 cm signal emitted during these cosmic epochs might provide
a viable tool to discriminate the two scenarios.

The 21 cm brightness temperature, relative to the CMB, is given by

\be
\delta T_b & = & 27 x_\mathrm{HI} (1+\delta) \left(\frac{\Omega_b h^2}{0.023}\right) 
\left(\frac{0.15}{\Omega_m h^2} \frac{1+z}{10}\right)^{1/2}  \nonumber \\
& & \times \left(\frac{T_S-T_\gamma}{T_S}\right) \mathrm{mK},
\label{eq:tbright}
\ee
\noindent
where $\delta$ is the fractional overdensity, $x_\mathrm{HI}$ is the hydrogen 
neutral fraction, and $T_\gamma$ is the CMB temperature at redshift $z$ 
(see Furlanetto, Oh \& Briggs 2006~for a thorough review of the subject).  
The spin temperature $T_S$, which represents the excitation temperature of
the 21 cm transition, determines whether the signal will appear in emission 
(if $T_S > T_\gamma$) or in absorption (if $T_S < T_\gamma$). The evolution of
the spin temperature is controlled by three competitive processes: (i) 
absorption and stimulated emission of CMB photons, which tend to couple 
$T_S$ to the CMB temperature; (ii) collisions with neutral hydrogen atoms and free
electrons, which tend to couple $T_S$ to the kinetic temperature of
the gas $T_K$; (iii) the absorption and subsequent spontaneous emission of Ly$\alpha$
photons, which mix the hyperfine levels and couples $T_S$ to an effective colour
temperature $T_C$ which, in most situations, can be well approximated by $T_K$. 
The latter process is referred to as Wouthuysen-Field (W-F) mechanism and it is effective in 
the presence of substantial UV light, requiring about 1 Ly$\alpha$ photon per baryon 
\citep{Wouthuysen1952, Field1959}. 
As a result, the spin temperature evolution can be written as

\be
T_S^{-1} = \frac{T_\gamma^{-1} + x_C T_K^{-1} + x_\alpha T_C^{-1}}{1+x_C+x_\alpha},
\label{eq:tspin}
\ee
\noindent
where
\[
x_C \equiv \frac{T_{21}}{T_\gamma} \frac{R_{21}^C}{A_{21}} \qquad   \mathrm{and} \qquad   
x_\alpha \equiv \frac{T_{21}}{T_\gamma} \frac{R_{21}^\alpha}{A_{21}}
\]
\noindent
are the coupling coefficients for collisions and UV scattering, $T_{21} = 0.068$~K and 
$A_{21} = 2.85 \times 10^{-15}\mathrm{s}^{-1}$ are the equivalent temperature and 
spontaneous emission rate of the 21 cm transition, and 
$R_{21}^C = n_\mathrm{H} \, k_{21}^\mathrm{H} + n_\mathrm{e} \, k_{21}^\mathrm{e}$
($R_{21}^\alpha$) is the rate coefficient for spin de-excitations by H-H and H-e 
collisions (UV scattering). In this analysis, we have used the value of 
$k_{21}^\mathrm{H}$ tabulated in \citet{Zygelman2005}\footnote{For kinetic temperatures in the
interval 1~K$<T_K<$300~K we have used the recommended rates tabulated in column (4) of 
Table 2 and, for $T> 300$~K we have applied the suggested analytic fit.}, and the 
functional form of $k_{21}^\mathrm{e}(T_K)$ reported in \citet{Liszt2001}. The rate coefficient
for the W-F mechanism is given by \citep{Hirata2006} 

\[
R_{21}^\alpha = \frac{8}{9} \pi \lambda_\alpha^2 \gamma S_\alpha J_\alpha,
\]
where $\lambda_\alpha = 1216$~\AA, $\gamma = 50$MHz is the half width at half-maximum of the
Ly$\alpha$ resonance, $J_\alpha$ is the flux of Ly$\alpha$ photons (in cm$^{-2}$s$^{-1}$Hz$^{-1}$sr$^{-1}$),
and $S_\alpha$ is a factor of order unity that accounts for spectral distortions. We have used the
numerical fits of \citet{Hirata2006} for $S_\alpha$ and the effective colour temperature $T_C$ which appears
in equation~(\ref{eq:tspin}).

Using equations (\ref{eq:tbright}) and (\ref{eq:tspin}), we can compute the all-sky 21 cm background 
signals predicted by the two radiative feedback models. We therefore take the redshift evolution 
of the gas kinetic temperature, $T_K$, and neutral hydrogen fraction, $x_\mathrm{HI}$, for model
CF06 and G00, shown respectively in panels (i) and (a) of Figs.~\ref{fig:modelCF06} and \ref{fig:modelG00}. 
We compute the Ly$\alpha$ background as

\[
J_\alpha (z) = \frac{(1+z)^3}{4 \pi} \int_z^\infty dz^\prime \frac{dl}{dz^\prime} 
 \epsilon({\nu,z^\prime}) \,  e^{-\tau_\mathrm{eff}(\nu_\alpha,z,z^\prime)},
\]

\noindent
where $\nu = \nu_{\alpha} (1+z^\prime)/(1+z)$, $dl/dz$ is the proper line element, and 
$\tau_\mathrm{eff}(\nu_\alpha,z,z^\prime)$ is the effective optical depth of the IGM
to radiation emitted at $z^\prime$ and observed at $z$ at frequency 
$\nu_\alpha$ (see section 2.2 of Salvaterra \& Ferrara 2003 for a full 
description of the IGM modelling).
The comoving emissivity $\epsilon(\nu,z)$ is computed as

\[
\epsilon(\nu,z)=\int_z^\infty dz^\prime l_\nu(t_{z,z^\prime})\dot{\rho}(z^\prime),
\]

\noindent
where $l_\nu(t_{z,z^\prime})$ is the template specific luminosity for a 
stellar
population of age $t_{z,z^\prime}$ (time elapsed between redshift $z^\prime$
and $z$). The final value of $\epsilon(\nu,z)$ is then computed by summing over
the Pop II and Pop III contribution (Salvaterra et al. 2006).
\begin{figure}
\center{ \epsfig{file=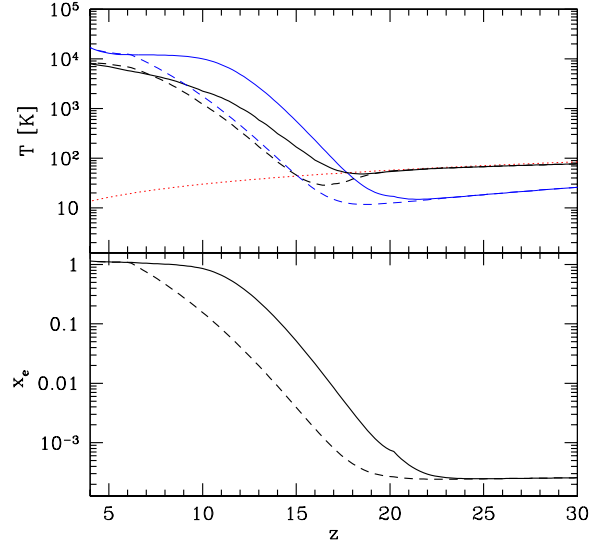, width=8.5cm} }
\caption{{\it Top panel:} redshift evolution of the spin (thick lines) and gas kinetic 
(thin lines) temperatures 
predicted by the two models. Solid lines refer to model CF06; dashed lines to model G00. 
For comparison, we also show the evolution of the CMB temperature (dotted line).
{\it Bottom panel:} corresponding evolution of the free electron fraction.}
\label{fig:tspin}
\end{figure}
In Fig.~\ref{fig:tspin} we show the evolution of the spin temperature, gas kinetic temperature
and CMB temperature for the two models. As expected, at very high redshifts the spin temperature
is coupled to the CMB, because collisions are negligible, and it decouples from the CMB because of 
the W-F effect, which determines the evolution at $z \lsim 20$. As the star formation rate is 
dominated by Pop II stars, the Ly$\alpha$ background is essentialy maintained by 
Pop II stars at all redshifts, and it is almost independent of the small halos  hosting Pop III stars, 
which are heavily affected by radiative feedback. As a consequence, the spin temperature
is driven towards the corresponding color (kinetic) temperature with comparable rates in the two
models; the different behavior in the redshift range $7 \lsim z \lsim 19$ is all due to the
dissimilar color (kinetic) temperature evolution caused by the two radiative    feedback models.
In particular, in model G00 the gas kinetic temperature is heated above the CMB value only at 
$z \lsim 15$. Therefore, between $15 \lsim z \lsim 20$, $T_S < T\gamma$ and we expect to see 
an absorption feature in the 21 cm signal which should be almost negligible in model CF06.
Note that X-rays from supernovae, X-ray binaries and mini-QSOs can be an important heating
agent for the neutral IGM. To cancel the predicted absorption feature, X-ray heating should
increase the kinetic temperature from $\sim 10$~K to values above $\sim 30-40$~K in the redshift
(time) interval $17 < z < 20$ ($\sim 50$~Myr). However, assuming that a fraction of 0.01 (0.1) of the 
total energy at Ly$\alpha$ is emitted in X-rays (Chen \& Miralda-Escud\'e 2004), the temperature
evolution of neutral regions predicted by the CF06 and G00 models is foud to deviate from adiabatic 
cooling only at $z < 11$ ($< 13$).

\begin{figure}
\center{ \epsfig{file=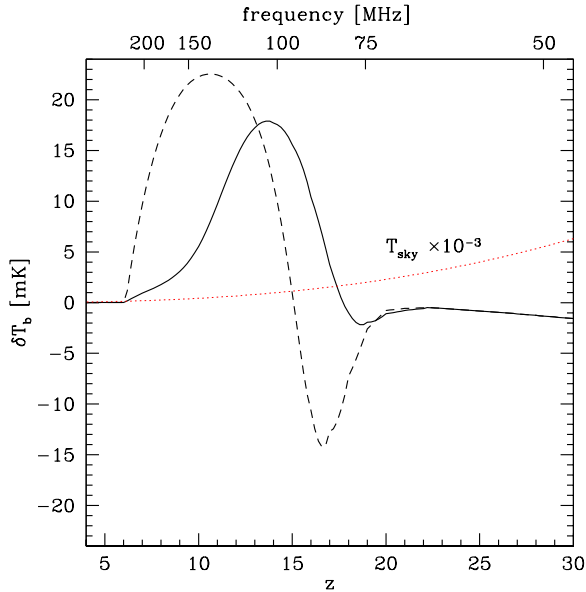, width=8.5cm} }
\caption{Predicted all-sky 21 cm brightness temperature relative to the CMB. Solid line
refers to model CF06, dashed line to model G00. The upper x-axis shows a reference 
observational frequency scale. The dotted line shows an estimate of the foreground
signal at these frequencies, rescaled by a factor $10^{-3}$ (see text).}
\label{fig:tbright}
\end{figure}

Fig.\ref{fig:tbright} shows the predicted 21 cm brightness temperature as a function of redshift.
The upper x-axis shows some reference observational frequency values, computed as $\nu=\nu_{21}/(1+z)$, 
where $\nu_{21}=1420$~MHz.
The expected absorption
signal in model G00 appears in the frequency range 75 - 100 MHz and has an amplitude of $\approx 15$~mK; 
this feature is far less evident in model CF06. At higher frequencies, the late reionization 
history predicted in model G00 (see the bottom panel of Fig.~\ref{fig:tspin}) causes a shift in the 
emission signal, which is larger than for model CF06. Note that since this is an all-sky signal,
single-dish observations from existing and planned low-frequency radio-telescopes can reach the
required mK sensitivity. The largest limitation to these measurements is the presence of strong
foregrounds in the relevant frequency range. The dotted line is a naive estimate of the 
contamination of a quiet region of the sky from foregrounds (mainly Galactic synchrotron), 
taken from \citet{Furlanettoreview2006}. 
It implies that the predicted signals should be completely
screened by foregrounds, which are three orders of magnitude larger. 

\begin{figure}
\center{ \epsfig{file=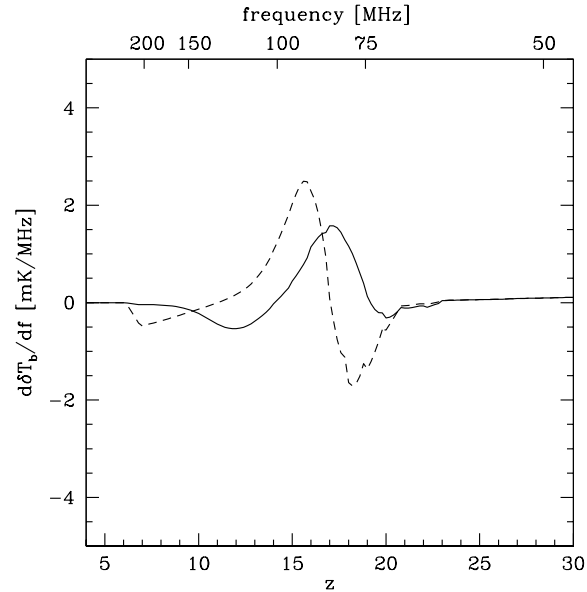, width=8.5cm} }
\caption{Frequency gradient of the 21 cm brightness shown in Fig.~\ref{fig:tbright}, adopting
the same line coding.}
\label{fig:tgrad}
\end{figure}

Since the frequency dependence of the signals and foregrounds are different, 
the gradient of the brightness fluctuation with frequency shows some
spectral features that might help to discriminate the signal from the 
relatively smooth foreground. Fig.~\ref{fig:tgrad} shows the predicted gradient
for the two models. The largest differences are due to the deeper absorption 
feature present in model G00, and to the shift of the emission preceding 
reionization. Clearly, these are of order $\approx 1-2$~mK/MHz and have to be extracted
from a $|dT_\mathrm{sky}/d\nu| \gsim 3$~K/MHz foreground so that their detection
will be very challenging. A discussion of possible strategies for this high-precision
21 cm measurement can be found in \citet{Furlanettoreview2006}.    

\begin{figure}
\center{ \epsfig{file=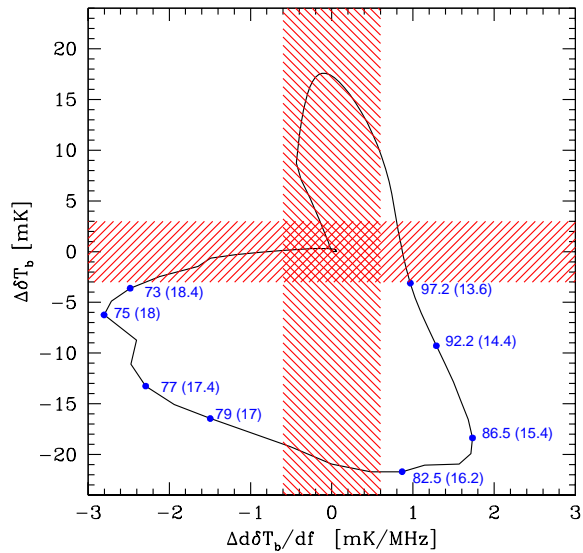, width=8.5cm} }
\caption{Difference in the brightness temperature and gradients between the two models (see text). The
shaded areas represent the regions where the signals will not be distinguishable given the foreseen 
sensitivity of future 21 cm experiments. Numbers along the curves indicate the frequency in MHz (and
the corresponding redshift) at which the signals would be observed.}
\label{fig:obs}
\end{figure}

Following \citet{Valdes2007}, we can tentatively assume that a successful 
detection requires the difference in the brightness temperature among the two 
models to be 
\[
\Delta \delta T_b = (\delta T_b)_{\rm{G00}} - (\delta T_b)_{\rm{CF06}} > 3~\mbox{mK} 
\]
and the difference in the gradients to be,
\[
\Delta \frac{d\delta T_b}{df} = \left(\frac{d\delta T_b}{df}\right)_{\rm{G00}} - \left(\frac{d\delta T_b}{df}\right)_{\rm{CF06}} > 0.6~\mbox{mK/MHz}.
\]
These conditions, based on the foreseen sensitivity of future 21 cm experiments, should take into
consideration the difficulties associated with foregrounds removal. Using these limits as a guideline,
Fig.\ref{fig:obs} shows that the two radiative feedback models could be discriminated through the differences 
in their predicted 21 cm background signals in the observed frequency ranges 73-79 MHz and 82.5-97.2 MHz,
which correspond to the redshift intervals 17-18.4 and 13.6-16.2, respectively.    

\section{Conclusions}
\label{summary}
Using a self-consistent semi-analytical model developed by \citet{Choudhury2005} and
recently updated by \citet{Choudhury2006}, we have explored the effect of reionization
and its associated radiative feedback on galaxy formation. The suppression of star formation in 
low-mass galaxies due to the increase in temperature of the cosmic gas in ionized regions has 
been modelled according to two different feedback prescriptions: 
(i) galaxies can form stars unimpeded provided
that their circular velocity is larger than a critical threshold, which is not fixed
to a constant value but evolves according to gas temperature (model CF06); 
(ii) depending on the mass of the galaxy, the fraction of gas available to star formation is 
reduced with respect to the universal value and it is fully specified by the filtering
mass at that redshift (model G00). We then constrain the four free-parameters
of each model with existing observational data, such as the redshift evolution of Lyman-limit 
absorption systems, the Gunn-Peterson and electron scattering optical depths, the cosmic star 
formation history, the number counts of high-$z$ sources in the NICMOS HUDF, and thermal
history of the IGM. We find that:

\begin{enumerate}
\item The two models provide equally good fit to existing observations, which therefore
are not able to break the degeneracy among different radiative feedback implementations.
\item Given that Pop III stars are preferentially hosted in the smallest galaxies at each
redshift, the largest difference between the two models reside in the predicted Pop III
star formation history: in models CF06, Pop III stars are fully expired by $z \sim 6$, whereas
in model G00 Pop III stars continue to form at lower $z$, though with decreasing rate. 
\item The reionization and thermal history of the IGM are very different: reionization starts 
at $z \lsim 15-20$, and it is already 85\% complete by $z \sim 10$ in model CF06, while, at 
the same $z$, the ionized fraction is only $16\%$ in model G00. 
Both models match SDSS constraints on the evolution of the neutral hydrogen fraction
at $z < 7$, but predict different Thomson optical depth, with $\tau_\mathrm{e}=0.1017, 0.0631$, for
model CF06 and G00, respectively, in agreement with {\it WMAP} 3-yr data. In principle, one
can use this property to distinguish between the two models, particularly if the observed value
of $\tau_e$ is better constrained by future CMB experiments like PLANCK.
\item Given the different gas ionization fraction and temperature evolution in the range 
$7 \lsim z \lsim 20$,  the two models predict different global 21 cm background signals in the observed
frequency range $75 \mathrm{MHz} \lsim \nu \lsim 200$~MHz. The largest differences in the two models
are represented by a $\sim 15$~mK absorption feature in the range 75-100 MHz in model G00
(which is nearly absent in model CF06), and by a global shift of the emission feature preceding
reionization towards larger frequencies in the same model. Single dish observations with
existing or forthcoming low-frequency radio telescopes such as LOFAR, 21CMA, MWA, and SKA 
can achieve mK sensitivity allowing the identification of these signals provided that foregrounds, 
which are expected to be three orders of magnitude larger, can be accurately subtracted.
\item The best observational frequencies to discriminate the radiative feedback models through their 21 cm
background signal are 73-79 MHz and 82.5-97.2 MHz, where the expected differences in brightness 
temperatures and gradients are large enough to be detectable with future 21 cm experiments.
\end{enumerate}

How robust is the proposed method of differentiating the two radiative
feedback models? It is important to stress that, for a given radiative feedback
model, it is not possible to find a different set of parameters which provides 
an equally good fit to the data presented in figures ~\ref{fig:modelCF06} 
and \ref{fig:modelG00} and yet produce a different 21cm signal. This is 
because the main difference between the G00 and CF06 models depends on 
(i) the confinement of Pop III stars in the smallest star forming halos 
by chemical feedback, and on (ii) how radiative feedback affects the gas content
and therefore the star formation efficiency in these small mass objects. 
For the same reasons, we do not expect the resulting 21cm signals to be 
affected by varying star formation efficiencies and escape fractions with 
redshift and/or mass: at $5 < z < 20$ more than 75\% of Pop III halos are 
confined in a 0.5 dex mass bin centered on the minimum mass to form stars. 
Therefore we believe that the differences we predict in the 21cm signals 
are robust. 

The absorption and emission features that we find are comparable to the 21 cm signatures
expected in models which consider the high-redshift signals emerging from the Dark Ages, 
following the appearence of first luminous sources \citep{Chen2004, 
Sethi2005, Furlanetto2006}, or induced by decaying/annihilating dark matter 
\citep{Valdes2007}. It is clear that measuring the 21 cm background would offer 
valuable insights into these early cosmic epochs. Our analysis suggests that 
self-consistent reionization models which are compatible with a large set of
existing observational data predict different 21 cm background signals, which
reflect how the process of star formation in the smallest galaxies is affected
by radiative    feedback. Future 21 cm data, complemented by the available
observational data from SDSS, WMAP, thermal history of the IGM, high-$z$
number counts, and the cosmic star formation history, might break current 
degeneracies and constrain the reionization process and its sources over
the redshift range $6 < z < 20$.  

\section*{Acknowledgments}
We are grateful to Benedetta Ciardi for profitable discussions and to DAVID 
members\footnote{http://www.arcetri.astro.it/science/cosmology} for fruitful
comments. CB acknowledges the support by the ASI contract 
``Planck LFI Activity of Phase E2''.

\label{lastpage}
\end{document}